\documentclass[journal]{IEEEtran}
\usepackage{amsfonts}  
\usepackage{color}

\usepackage{cite}

\usepackage[dvips]{graphicx}

\usepackage{amsmath} 
\usepackage{bm}
\newtheorem{theorem}{\textbf{Theorem}}

\usepackage{algorithm}
\usepackage{algorithmic}

\usepackage{array}

\usepackage[caption=false,font=footnotesize]{subfig}

\begin{document}

\title{Optimum Fairness for Non-Orthogonal \\ Multiple Access}

\author{Ting~Qi,
        Wei~Feng,~\IEEEmembership{Member,~IEEE,}
        Yunfei Chen,~\IEEEmembership{Senior Member,~IEEE,}
        Youzheng~Wang,~\IEEEmembership{Member,~IEEE}
        \thanks{This work was partially supported by Natural Science Foundation (91638205, 91438206, 61621091).}
        \thanks{T. Qi and Y. Wang are with Tsinghua Space Center, Tsinghua University, Beijing 100084, P. R. China (e-mail: qit13@mails.tsinghua.edu.cn; yzhwang@mail.tsinghua.edu.cn). W. Feng is with the Tsinghua National Laboratory for Information Science and Technology, Department of Electronic Engineering, Tsinghua University, Beijing 100084, P. R. China (e-mail: fengwei@tsinghua.edu.cn). Y. Chen is with the School of Engineering, University of Warwick, Coventry CV4 7AL, United Kingdom (e-mail: Yunfei.Chen@warwick.ac.uk).}
        }

\maketitle

\begin{abstract}
This paper focuses on the fairness issue in non-orthogonal multiple access (NOMA) and investigates the optimization problem that maximizes the worst user’s achievable rate. Unlike previous studies, we derive a closed-form expression for the optimal value and solution, which are related to Perron-Frobenius eigenvalue and eigenvector of a defined positive matrix. On this basis, we propose an iterative algorithm to compute the optimal solution, which has linear convergence and requires only about half iterations of the classical bisection method.
\end{abstract}


\IEEEpeerreviewmaketitle

\section{Introduction}
\IEEEPARstart{T}{he} next generation wireless communication networks will face the challenges of rapid growth of data traffic. Moreover, the Internet of Things (IoT) becomes a new communication paradigm that enables anyone and anything to be served at anytime and anyplace, which bring the challenge of massive connectivity.
Non-orthogonal multiple access (NOMA) has been widely recognized as a promising technology to address these challenges for its potential to enhance the spectral efficiency \cite{Ding2017}. By letting multiple users share the limited radio spectrum, more users than the number of orthogonal resources are able to be simultaneously served in NOMA \cite{Ding2015a,Qi2017}. At the receiver, advanced multiuser detection, such as the successive interference cancellation (SIC) and the message passing algorithm, is used to distinguish the mutual interfering users \cite{Qi2017a}. In NOMA, in addition to spectral efficiency of the system, user fairness is also an important issue, especially for homogeneous networks, where the traffic arrival rate is almost identical.

Power allocation is crucial to the fairness of NOMA because the rates of users are closely related due to the multiple access interference. For NOMA with practical modulation of finite constellation sizes, the total mutual information evaluated by Monte Carlo simulations was maximized when allocating the limited power in \cite{Choi2016a}.
In the downlink multiuser systems, the sum-capacity achievable power allocation method allocates all power to the user with the highest channel gain \cite{Choi2016}. This scheme is unfair to the users with worse channel conditions, who may not be served over a period of time and suffer long latency of traffic. Therefore, it is important to guarantee fairness when performing power allocation.

In \cite{Oviedo2016}, a fair NOMA scheme was introduced by guaranteeing users to achieve the rate at least as good as orthogonal multiple access (OMA). For NOMA systems with multiple antennas, fairness can be realized by transmit antenna selection while improving the sum rate \cite{Uesuenbas2017}. The work in \cite{Timotheou2015} studied the max-min fairness with instantaneous channel state information (CSI) and min-max fairness with average CSI, respectively and developed a low-complexity bisection method to yield the optimal solution. The max-min fairness problem was further considered in NOMA based cognitive radio networks \cite{Zeng2017}.

For fading channel with only statistical CSI at the transmitter, various fair power allocation schemes for NOMA were investigated in \cite{Cui2016,Shi2016}. By constrains on the outage probability, the problem of maximizing the minimum outage rate was studied \cite{Cui2016}. The authors in \cite{Shi2016} considered the minimum weighted success probability maximization problem with power allocation, decoding order and user grouping being taken into account.

In this paper, we formulate the power allocation problem that maximizes the minimum achievable rate among users. The novelty of this work is that by transforming the optimal condition to an eigen equation, we derive the closed-form optimal value and solution, which are functions of Perron-Frobenius (PF) eigenvalue and
eigenvector of a defined matrix.  Accordingly, a low-complexity iterative algorithm is proposed to compute the optimal solution. The proposed algorithm is proved to have linear convergence, and achieves the same accuracy twice as fast as the bisection method. Moreover, the minimum user rates and Jain's fairness index of different power allocation schemes in NOMA and OMA are compared by simulation.


\section{System Model and problem formulation}

Consider the downlink multiuser transmission where one base station (BS) sends separate information to multiple users. All terminals are equipped with one antenna. In the power-domain NOMA scheme, the BS transmits the linear superposed signals of the $K$ users using the same resource block and the users perform multiuser detection, i.e., SIC, to extract their own information.

Let $x_k$ be the symbol transmitted to user $k$ with a normalized average power of 1, i.e., $\mathbb{E}[|x_k|^2] = 1, \forall k\in \mathcal{K}=\{1,\ldots,K\}$. The channel coefficient from the BS to user $k$ is denoted by $h_k$. All channels exhibit independent block fading and remain constant during the block, but change independently across different blocks. Assume that the BS has perfect CSI of each user. The signal received by user $k$ during one block can be written as
\begin{equation}\label{eq.RxModel}
y_k = h_k \sum_{k=1}^{K} \sqrt{P_k} x_k + w_k, \quad k \in \mathcal{K},
\end{equation}
where $w_k\sim\mathbb{C}\mathcal{N}(0,1)$ is i.i.d complex additive white Gaussian noise, and $P_k$ is the power allocated to user $k$.

Without loss of generality, assume that the channels are sorted as $|h_1|>|h_2|>\cdots >|h_K|$, i.e., user $k$ always holds the $k$-th strongest instantaneous channel. The achievable rate of user $k$ is given as
\begin{equation}\label{eq.rate}
R_k = \left\{
\begin{array}{l}
\log_2 \left(1+ P_1 |h_1|^2\right), \; \mathrm{if}\, k=1, \\
\log_2 \left(1+ \frac{P_k |h_k|^2}{|h_k|^2 \sum_{j=1}^{k-1} P_j +1}\right), \; \mathrm{if}\, k \in \mathcal{K}\setminus \{1\}. \\
\end{array} \right.
\end{equation}

In NOMA, although multiple users can be simultaneously served, users with relatively weak channels suffer from more severe MAI than those with strong channels. It is necessary to guarantee the user fairness when allocating the limited power resources.
Therefore, we formulate the max-min fairness problem of maximizing the minimum achievable rate, given as
\begin{subequations}\label{eq.problem}
\begin{align}
\label{eq.obj}
& \mathop{\mathrm{max}}\limits_{\bm{P}} \; \mathop{\mathrm{min}}\limits_{k \in \mathcal{K}} \;  R_k \\
\label{eq.cons1}
& \mathrm{s.t.} \quad \sum_{k=1}^K P_k \leq P_T, \\
\label{eq.cons2}
& \phantom{\mathrm{s.t.} \quad} P_k \geq 0, \; \forall k \in \mathcal{K},
\end{align}
\end{subequations}
where $\bm{P}=[P_1,\ldots, P_K]^\mathrm{T}$, $P_T$ is the total transmission power. The optimal value of problem (\ref{eq.problem}) is denoted as the fairness rate.

\section{Closed-form Solution}
Problem (\ref{eq.problem}) is not convex and hence is hard to solve directly using standard convex optimization tools.
In \cite{Timotheou2015,Zeng2017}, the authors proved that the problem is quasi-concave and developed a bisection method to solve a sequence of linear programs to obtain the optimal solution. Different from that work, by analyzing the optimum condition and changing the expression at optimality, we derive the closed-form optimal solution and value for problem (\ref{eq.problem}) leveraging the Perron-Frobenius theorem. The following theorem states the results.

\begin{theorem}\label{thm.solution}
The optimal value and solution of problem (\ref{eq.problem}), denoted by $R^*$ and $\bm{P}^*$, respectively, are given by
\begin{align}\label{eq.optimal}
R^* &= \log_2\left( 1+\frac{1}{\lambda_{\mathrm{pf}}} \right), \\
\bm{P}^* &= \frac{\bm{v}}{\bm{1}^\mathrm{T}\bm{v}} P_T,
\end{align}
where $\lambda_{\mathrm{pf}}$ and $\bm{v}$ are Perron-Frobenius (PF) eigenvalue and eigenvector of the positive matrix defined by $\bm{A}+\bm{b1}^\mathrm{T}$, respectively. Here
\begin{equation}\label{eq.Ab1}
\bm{A} = \left(
      \begin{array}{cccccc}
        0 & 0 & 0 & \cdots & 0 & 0 \\
        1 & 0 & 0 & \cdots & 0 & 0 \\
        1 & 1 & 0 & \cdots & 0 & 0 \\
        1 & 1 & 1 & \cdots & 0 & 0 \\
          &   &   & \ddots &   &   \\
        1 & 1 & 1 & \cdots & 1 & 0 \\
      \end{array}
    \right),
\bm{b} = \left(
           \begin{array}{c}
             \frac{1}{P_T|h_1|^2} \\
             \frac{1}{P_T|h_2|^2} \\
             \vdots \\
             \frac{1}{P_T|h_K|^2} \\
           \end{array}
         \right)
\end{equation}
are $K\times K$ matrix and $K\times 1$ vector, respectively, and $\bm{1}$ represents $K\times 1$ vector of all ones.
\end{theorem}

\begin{IEEEproof}
We first prove that the optimal solution to problem (\ref{eq.problem}) is achieved when all user rates are equal and the constraint of (\ref{eq.cons1}) is tight at optimality.

Denote the minimum rate as $R_{\mathrm{min}}$. Assume that there exist user $j$ and $k$ with rate $R_j>R_{\mathrm{min}}$ and $R_k=R_{\mathrm{min}}$, respectively. Then $R_{\mathrm{min}}$ can be maximally improved by reducing $P_j$ and increasing $P_k$ such that $R_j=R_k$. Thus, all users have equal rates at optimality.
Suppose $\sum_{k=1}^K P_k < P_T$ at optimality. The objective function in (\ref{eq.obj}) can be strictly improved by increasing the power of all users proportionally such that $\sum_{k=1}^K P_k = P_T$, since $R_k(\bm{P})$ for all $k$ increases monotonically. This contradicts the assumption, thus $\sum_{k=1}^K P_k = P_T$ is true at optimality.

Since all user rates are equal to $R^*$ at optimality, substituting $\bm{P}^*$ into (\ref{eq.rate}), we have
\begin{align}\label{eq.optimal.rate}
\begin{split}
\frac{1}{2^{R^*}-1}P_1^* &= \frac{1}{|h_1|^2},\\
\frac{1}{2^{R^*}-1}P_k^* &= \sum_{j=1}^{k-1}P_j^* +\frac{1}{|h_k|^2}, \, \forall k \in \mathcal{K}\setminus \{1\}.
\end{split}
\end{align}

Divide both sides of the $K$ equations (\ref{eq.optimal.rate}) by $P_T$ and let $\bm{s}^* = \bm{P}^*/P_T$. Then equations (\ref{eq.optimal.rate}) can be rewritten as
\begin{equation}\label{eq.fixed.point}
\frac{1}{2^{R^*}-1}\bm{s}^* = \bm{A}\bm{s}^* +\bm{b}.
\end{equation}

Noting that $\sum_{k=1}^K P_k^* = P_T$, we have $\sum_k s_k^*=1$, i.e., $\bm{1}^\mathrm{T}\bm{s}^*=1$. Then, equation (\ref{eq.fixed.point}) can be transformed to
\begin{equation}\label{eq.fixed.point1}
\frac{1}{2^{R^*}-1}\bm{s}^* = (\bm{A}+\bm{b1}^\mathrm{T})\bm{s}^*.
\end{equation}
Let $\bm{B}= \bm{A}+\bm{b1}^\mathrm{T}$ and the equation (\ref{eq.fixed.point1}) can be seen as the eigen equation of $\bm{B}$.
It is easy to verify that all entries of matrix $\bm{B}$ are positive. According to the Perron-Frobenius theorem \cite{Blondel2005}, $\bm{B}$ has unique largest real positive eigenvalue, denoted by PF eigenvalue $\lambda_{\mathrm{pf}}$, and the associated PF eigenvector can have all positive elements.
Therefore, $R^*$ can be solved by $\frac{1}{2^{R^*}-1}=\lambda_\mathrm{pf}$ and $\bm{s}^* =\frac{\bm{v}}{\bm{1}^\mathrm{T}\bm{v}}$ is  the 1-norm normalized eigenvalue. This completes the proof.
\end{IEEEproof}

The optimal solution given in Theorem \ref{thm.solution} can be numerically computed by the algorithm proposed in section IV. To analyze the influence of total power and channel gains, we evaluate $\lambda_{\mathrm{pf}}$ by the following upper and lower bound, from which some insights can be observed.  Since $\bm{B}$ is positive, the spectral radius, equal to $\lambda_{\mathrm{pf}}$, is bounded by \cite{Minc1988}
\begin{equation}\label{eq.rho_bound}
\mathop{\mathrm{min}}\limits_{i} \,  \sum_{j=1}^K b_{ij} \leq \lambda_{\mathrm{pf}} \leq \mathop{\mathrm{max}}\limits_{i} \,  \sum_{j=1}^K b_{ij},
\end{equation}
where $b_{ij}$ is the entry of the $i$-th row and $j$-th column of matrix $\bm{B}$.
Substituting (\ref{eq.Ab1}) into (\ref{eq.rho_bound}), we have
\begin{equation}\label{eq.rho_bound1}
\frac{1}{P_T}\sum_{k=1}^K \frac{1}{|h_k|^2} \leq \lambda_{\mathrm{pf}} \leq K-1+\frac{1}{P_T}\sum_{k=1}^K \frac{1}{|h_k|^2}.
\end{equation}
Therefore, $R^*$ is bounded by
\begin{align}\label{eq.R_bound}
\begin{split}
 &R^* \geq \log_2 \left( 1\!+\! \frac{1}{ K\!-\!1 \!+ \!\frac{1}{P_T}\!\sum_{k=1}^K \!\frac{1}{|h_k|^2}} \right), \\
 &R^*  \leq  \log_2 \left( 1\!+\! \frac{P_T}{\sum_{k=1}^K \!\frac{1}{|h_k|^2}} \right),
\end{split}
\end{align}
from which we can observe that, for a fixed $K$ and channel gains, $R^*$ is a logarithmic function of $P_T$.

\section{Algorithm with linear convergence}
In this section, we propose an iterative algorithm with linear convergence to compute the optimal solution given in Theorem \ref{thm.solution}.

Normalizing the 1-norm of both sides of (\ref{eq.fixed.point1}), we find that $\bm{s}^*$ is the unique positive fixed point of the equation
\begin{equation}\label{eq.Bb}
\bm{s}= \frac{(\bm{A}+\bm{b1}^\mathrm{T})\bm{s}}{\|(\bm{A}+\bm{b1}^\mathrm{T})\bm{s}\|_1}.
\end{equation}
This inspires an iterative algorithm, summarized in Algorithm \ref{Alg.iterative}, which can find the optimal solution to problem (3).

\begin{algorithm}
\caption{The proposed iterative algorithm for solving problem (\ref{eq.problem})}
\label{Alg.iterative}
\begin{algorithmic}
\STATE \textbf{Input:} The total power $P_T$, channel gain $|h_k|, \, k \in \mathcal{K}$ and tolerance $\epsilon$.
\STATE \textbf{Output:} The optimal solution $\bm{P}^*$ and value $R^*$.
\STATE \textbf{Initialize} $i=1$, let $P_k^{(1)}=P_T/K, \, \forall k$, calculate rate $R_k^{(1)}, \, \forall k$ and find the minimum rate $R_{\mathrm{min}}^{(1)}$;
\REPEAT
\STATE $i=i+1$;
\STATE Update power $\bm{P}^{(i)}=(\bm{A}+\bm{b1}^\mathrm{T})\bm{P}^{(i-1)}$ with $\bm{A},\bm{b}$ given in (\ref{eq.Ab1});
\STATE Normalize power $\bm{P}^{(i)}= \frac{\bm{P}^{(i)}}{\|\bm{P}^{(i)}\|_1}P_T$;
\STATE Update rate $R_k^{(i)},\, \forall k$ and the minimum rate $R_{\mathrm{min}}^{(i)}$;
\UNTIL{$|R_{\mathrm{min}}^{(i)}-R_{\mathrm{min}}^{(i-1)}|<\epsilon$}
\STATE Set $\bm{P}^*=\bm{P}^{(i)},\, R^*=R_{\mathrm{min}}^{(i)}$.
\end{algorithmic}
\end{algorithm}

We then analyze the convergence rate of the proposed algorithm. Denote the $K$ eigenvalues of  $\bm{B}$ by $\lambda_1,\lambda_2,\ldots, \lambda_K$ in descending order of modulus.
Let $\bm{v}_1, \ldots, \bm{v}_K$ be the corresponding $K$ independent eigenvectors of $\bm{B}$. They form a basis of $\mathbb{R}^K$. Hence the initial vector $\bm{s}^{(1)}$ can be written as
\begin{equation}\label{eq.basis}
\bm{s}^{(1)}= a_1\bm{v}_1+a_2\bm{v}_2+\cdots +a_K\bm{v}_K,
\end{equation}
where $a_1,\ldots, a_K$ are scalars. Multiplying both sides of the equation by $\bm{B}^n$ yields
\begin{align}\label{eq.Bn}
\begin{split}
\bm{B}^n\bm{s}^{(1)} &= a_1\bm{B}^n\bm{v}_1+a_2\bm{B}^n\bm{v}_2+\cdots +a_K\bm{B}^n\bm{v}_K, \\
&= a_1\lambda_1^n \bm{v}_1 +a_2\lambda_2^n\bm{v}_2+\cdots +a_K\lambda_K^n\bm{v}_K, \\
& = a_1\lambda_1^n\left(\bm{v}_1+ \sum_{j=2}^K \frac{a_j}{a_1}\left(\frac{\lambda_j}{\lambda_1}\right)^n \bm{v}_j\right).
\end{split}
\end{align}

Note that $\bm{B}$ is positive and regular since $\bm{B}^n>0$ for $n\geq 1$. According to Perron-Frobenius theorem, we have $\lambda_1>|\lambda_2|\geq \cdots \geq |\lambda_K|$. Thus as $n\rightarrow \infty$, $\left(\frac{\lambda_j}{\lambda_1}\right)^n\rightarrow 0$ and $\bm{B}^n\bm{s}^{(1)}\rightarrow a_1\lambda_1^n \bm{v}_1$.  It is showed by (\ref{eq.Bn}) that $\bm{s}^{(n+1)}=\frac{\bm{B}^n\bm{s}^{(1)}}{\|\bm{B}^n\bm{s}^{(1)}\|_1}$ converges to $\bm{s}^*$ as fast as a geometric series. Therefore, the proposed algorithm has linear convergence since the error lies below a line on a log-linear plot of error versus iteration number \cite{Boyd2009}, as shown in the simulation.

Given a tolerance $\epsilon$, we can obtain that the required iteration is a linear function of $\log(\epsilon)$ from (\ref{eq.Bn}). In each iteration, there are $K$ computations of power and rate. Thus, the algorithm has a computational complexity of $\mathcal {O}\left(K\log(\epsilon)\right)$.

\section{Simulation Results and discussion}
\begin{figure}
  \centering
  \includegraphics[width=3.4in]{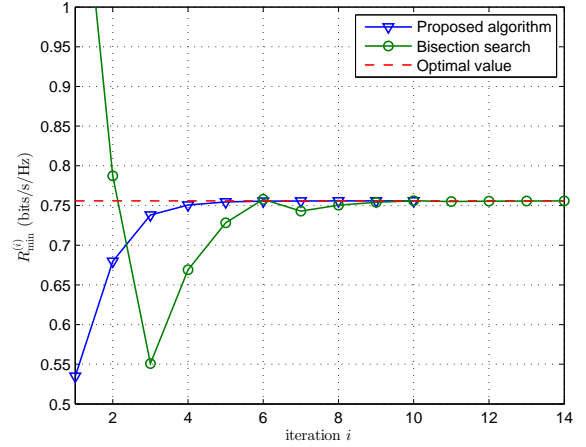}
  \caption{The convergence process with the iteration.}
  \label{fig.R_iterN}
\end{figure}

\begin{figure}
  \centering
  \includegraphics[width=3.4in]{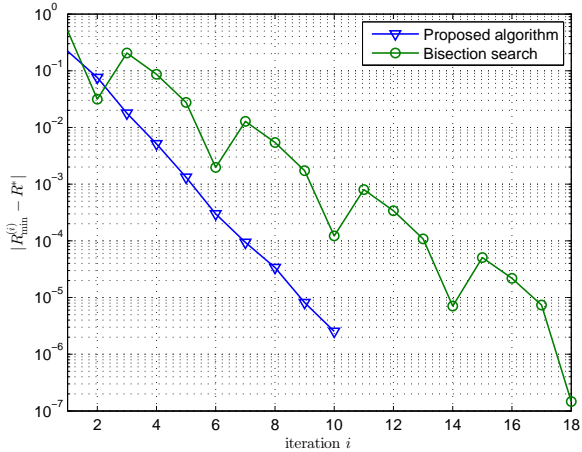}
  \caption{The error $|R_{\mathrm{min}}^{(i)}-R^*|$ versus iteration $i$.}
  \label{fig.logRerr_iterN}
\end{figure}

In this section, we present the simulation results for the proposed power allocation algorithm in NOMA.
We set the total transmission power $P_T=10 \mathrm{W}$, the number of user $K=4$ and the iteration tolerance $\epsilon=10^{-5}$. The minimum user rate in each iteration is depicted in Fig .\ref{fig.R_iterN}. The result for the bisection search method presented in \cite{Timotheou2015,Zeng2017} is also plotted for comparison. The bisection method starts with an interval containing the optimal value and determines whether the optimal value is in the upper or lower half of the interval in each iteration.  It can be seen that $R_{\mathrm{min}}^{(i)}$ of the proposed algorithm continually and quickly approaches the optimal value while the bisection method converges in an oscillating and slow way.
Fig. \ref{fig.logRerr_iterN} shows the error $|R_{\mathrm{min}}^{(i)}-R^*|$ versus iteration $i$. It verifies the convergence analysis that the proposed algorithm has linear convergence. Also, the proposed algorithm converges faster than the bisection method.

\begin{figure}
  \centering
  \includegraphics[width=3.4in]{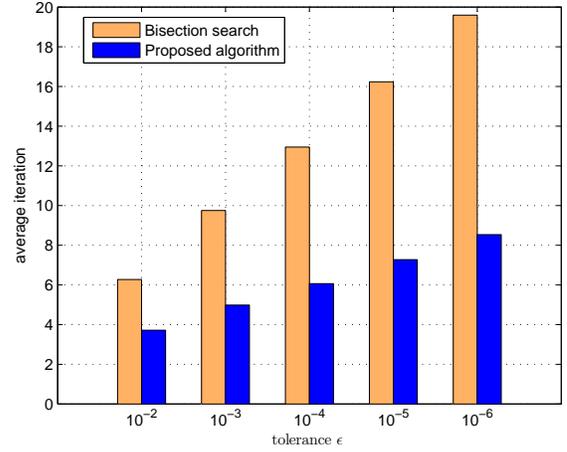}
  \caption{The average iteration versus the tolerance $\epsilon$.}
  \label{fig.average_iterN}
\end{figure}

\begin{figure}
  \centering
  \includegraphics[width=3.4in]{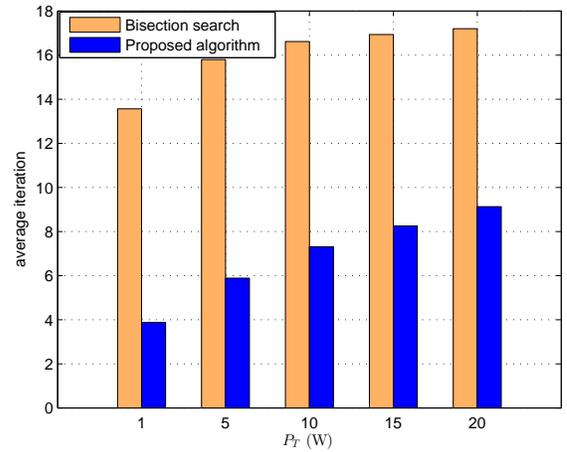}
  \caption{The average iteration versus the total power $P_T$.}
  \label{fig.average_Pt}
\end{figure}

Since the computational complexity is proportional to the number of iterations, we further compare the convergence rate. The average iteration evaluated by solving 1000 problems with randomly generated channel for different tolerance is presented in Fig. \ref{fig.average_iterN}. It shows that for identical tolerance, the proposed algorithm needs only half iterations of the bisection method. Moreover, Fig. \ref{fig.average_Pt} plots the average iteration versus $P_T$. It can be observed that the average iteration increases approximately linearly with $P_T$. The proposed algorithm reduces the iterations greatly and the performance gain is larger in the lower power regime. That is to say the proposed algorithm reduces the computation complexity by half as the complexity in each iteration is almost identical for the two algorithms.

\begin{figure}
  \centering
  \includegraphics[width=3.4in]{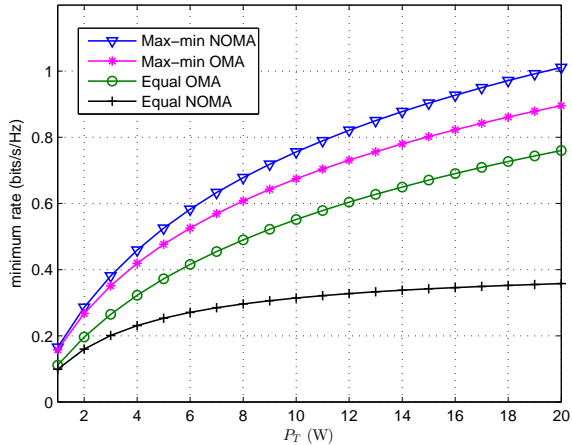}
  \caption{The minimum user rate of different power allocation scheme as the total power varies.}
  \label{fig.R_Pt}
\end{figure}

\begin{figure}
  \centering
  \includegraphics[width=3.4in]{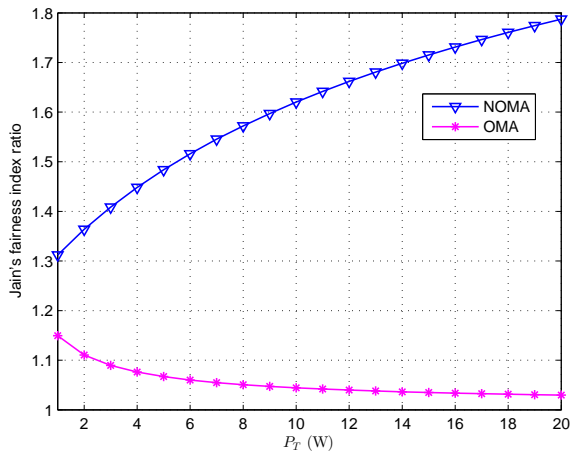}
  \caption{Jain's fairness index ratio between max-min and equal power allocation.}
  \label{fig.index_ratio}
\end{figure}

Fixing $K=4$ and randomly generating a set of channel gain $[|h_1|^2,|h_2|^2,\ldots,|h_K|^2]= [1.2389,0.7192,0.4322,0.3614]$, Fig. \ref{fig.R_Pt} presents the minimum rate of different power allocation scheme. The "Max-min NOMA" and "Max-min OMA" are the schemes that maximize the minimum user rate in NOMA and OMA, respectively, while "Equal NOMA/OMA" denotes equal power allocation among users. As analyzed in (\ref{eq.R_bound}), the fairness rate improves with $P_T$ logarithmically. The "Max-min NOMA" has higher fairness rate than the "Max-min OMA", and the max-min fair power allocation outperforms the equal power allocation in terms of guaranteeing the worst user's achievable rate. Moreover, it can be observed that fairness problem is more prominent in NOMA than OMA since the minimum user rate improves more dramatically from "Equal NOMA" to "Max-min NOMA". We use the Jain's fairness index, defined as
\begin{equation}\label{eq.index}
  F=\frac{(\sum_{k=1}^K R_k)^2}{K\sum_{k=1}^K R_k^2},
\end{equation}
to quantize the fairness of power allocation scheme. The "Max-min NOMA" and "Max-min OMA" achieve the best fairness with index 1. The fairness index of equal power allocation for NOMA and OMA can be computed using (\ref{eq.index}). Fig. \ref{fig.index_ratio} shows the fairness index ratio between max-min and equal power allocation for NOMA and OMA. It indicates that fairness is improved by max-min power allocation more significantly in NOMA than OMA. Furthermore, as $P_T$ increases, so does the fairness index ratio for NOMA, which implies that fairness issue becomes severer, while the opposite is true for OMA.

\section{Conclusion}
In this paper, we have formulated the power allocation problem for NOMA that maximizes the minimum achievable rate of users. By transforming the optimal condition to the eigen equation, we have derived the closed-form optimal solution based on the Perron-Frobenius theorem. Then an iterative algorithm with linear convergence has been proposed to obtain the optimal solution, which requires nearly half iterations of the bisection method. The minimum user rate is largely improved compared with the same scheme in OMA and equal power allocation.


\end{document}